\documentclass[12pt,preprint]{aastex}
\begin{document}
\newcommand{\lya}{Lyman-$\alpha$}
\newcommand{\eqw}{\hbox{EW}}
\def\erg{\hbox{erg}}
\def\cm{\hbox{cm}}
\def\sec{\hbox{s}}
\def\f17{f_{17}}
\def\Mpc{\hbox{Mpc}}
\def\cMpc{\hbox{cMpc}}
\def\pMpc{\hbox{pMpc}}
\def\Gpc{\hbox{Gpc}}
\def\nm{\hbox{nm}}
\def\km{\hbox{km}}
\def\kms{\hbox{km s$^{-1}$}}
\def\year{\hbox{yr}}
\def\Myr{\hbox{Myr}}
\def\Gyr{\hbox{Gyr}}
\def\deg{\hbox{deg}}
\def\arcsec{\hbox{arcsec}}
\def\microJy{\mu\hbox{Jy}}
\def\zre{z_r}
\def\fesc{f_{\rm esc}}
\def\lstar{\ifmmode {L_\star}\else
                ${L_\star}$\fi}
\def\phistar{\ifmmode {\phi_\star}\else
                ${\phi_\star}$\fi}

\def\ergcm2s{\ifmmode {\rm\,erg\,cm^{-2}\,s^{-1}}\else
                ${\rm\,ergs\,cm^{-2}\,s^{-1}}$\fi}
\def\ergsec{\ifmmode {\rm\,erg\,s^{-1}}\else
                ${\rm\,ergs\,s^{-1}}$\fi}
\def\kmsMpc{\ifmmode {\rm\,km\,s^{-1}\,Mpc^{-1}}\else
                ${\rm\,km\,s^{-1}\,Mpc^{-1}}$\fi}
\def\kpc{{\rm kpc}}
\def\taulya{\tau_{Ly\alpha}}
\def\taubar{\bar{\tau}_{Ly\alpha}}
\def\llya{L_{Ly\alpha}}
\def\ldlya{{\cal L}_{Ly\alpha}}
\def\nbar{\bar{n}}
\def\Msun{M_\odot}
\def\sqamin{\Box'}

\def\rss{R_{ss}}
\def\vss{V_{ss}}
\def\xvion{x_{V,\hbox{\scriptsize ion}}}
\def\xvn{x_{V,\hbox{\scriptsize n}}}
\def\ff{{\cal F}}

\title{The Volume Fraction of Ionized Intergalactic Gas at Redshift z=6.5}

\author{Sangeeta Malhotra \& James E. Rhoads}
\affil{Space Telescope Science Institute, 3700 San Martin Drive,
 Baltimore, MD 21218}
\email{san@stsci.edu,rhoads@stsci.edu}

\begin{abstract}

The observed number density of \lya\ sources implies a minimum volume of
the inter-galactic medium that must be ionized, in order to allow the
\lya\ photons to escape attenuation. We estimate this volume by assigning to
each \lya\ emitter the minimum Stromgren sphere that would allow half
its \lya\ photons to escape.  This implies a lower limit to ionized
gas volume fraction of 20-50\% at redshift z=6.5. This is a lower limit 
in two ways: First, we conservatively assume that the \lya\ sources seen
(at a relatively bright flux limit) are the only ones present; and second,
we assume the smallest Stromgren sphere volume that will allow
the photons to escape. 
This limit is completely independent of what ionizing photon sources
produced the bubbles.  Deeper \lya\ surveys are possible with present
technology, and can strengthen these limits by detecting a higher
density of \lya\ galaxies.

\end{abstract}

\keywords{galaxies: high-redshift,  (galaxies:) intergalactic medium}

\section{Introduction}

The epoch of reionization marks a phase transition in the universe,
when the intergalactic medium was ionized. Recent observations of
$z>6$ quasars show a Gunn-Peterson trough (Gunn \& Peterson 1965)
implying that the reionization of intergalactic
hydrogen was not complete until $z\approx 6$ (Becker et al 2001, Fan
et al 2002).  Yet microwave background observations imply substantial
ionization as early as $z\ga 15$ (Spergel et al 2003; Kogut et al
2003).  Transmitted flux seen in some quasar spectra also implies
inhomogenities in the ionized gas (Oh \& Furlanetto 2005; White et
al. 2005).  These results can be reconciled if reionization occurred
twice (e.g., Cen 2003), slowly (e.g. Gnedin 2004) or was substantially
inhomogenous (Malhotra et al. 2005, Oh \& Furlanetto 2005, Cen
2005). Information on the state of the inter-galactic gas--- the
ionized fraction, and spatial distribution of ionized gas--- is
needed, but is scant.

\paragraph{Lyman-$\alpha$ emitters as tests of ionized IGM}
\lya\ emitting galaxies provide another tool for probing reionization,
which is independent of and complementary to of the CMBR and
Gunn-Peterson diagnostics. \lya\ visibility tests offer a local probe
of neutral fractions $x_{HI}\sim 30 \%$ (Malhotra \& Rhoads 2004 (MR04)), while the Gunn-Peterson
trough saturates at $x_{HI}\sim 1\%$,(Fan et al 2002), and the CMBR
polarization provides an integral constraint of the ionized gas along
the line of sight.

Because \lya\ photons are resonantly scattered by neutral hydrogen,
\lya\ line fluxes are attenuated for sources in a significantly
neutral ($\langle x_{HI}\rangle \sim 0.1$) intergalactic medium (IGM)
(Miralda-Escud\'{e} 1998, Loeb \& Rybicki 1999, Haiman \& Spaans
1999).  To zeroth order this should produce a decrease in \lya\ galaxy
counts at redshifts beyond the end of hydrogen reionization (Rhoads \&
Malhotra 2001 (RM01)).  This does not mean that \lya\ emitters will
suddenly become invisible at some redshift, but rather that \lya\ flux
is strongly attenuated.  We consider three physical effects that
modify the attenuation of the \lya\ flux.

\paragraph{Effect of ionized bubbles}
Each galaxy creates a local bubble of ionized gas. If this is large
enough, \lya\ photons are redshifted by the time they reach the
neutral boundary and thus can escape.  Consider a galaxy in a
Stromgren sphere of radius $\rss$, surrounded by a fully neutral
IGM. The line center optical depth due to scattering by the damping
wing of neutral gas outside the galaxy's Stromgren sphere is given by
$\tau = 1.2 \pMpc / \rss$ (e.g., RM01 ).\footnote{In this paper we will write physical Mega-parsecs as
  \pMpc, and comoving Mega-parsecs as \cMpc. $1\pMpc = (1+z) \cMpc$.}
Thus, transmission becomes
significant when the physical radius of the HII region is $\ga 1.2
\pMpc$ (correpsonding to $\ga 9 \cMpc$ at redshift $z=6.5$),

For wavelengths separated from line center by $\Delta \lambda_{em}$, the
optical depth due to neutral IGM damping wing is
\begin{equation}
\tau = 1.2 \pMpc \times  \left(\rss + { \Delta \lambda_{em} \over
 \lambda_0} {c \over H} \right)^{-1}
\label{taueq}
\end{equation}
where $c$ is lightspeed and where the Hubble constant 
$H \approx 760 \kmsMpc \times [(1+z)/7.5]^{3/2}$ for 
a flat cosmology with $H_0 = 71 \kmsMpc$, $\Omega_m = 0.27$, 
and $\Omega_\Lambda = 0.73$ (see Spergel et al 2003).

The size of a typical Stromgren sphere that an isolated galaxy 
creates in the IGM can be directly related to its \lya\ luminosity
$\llya = 10^{43} L_{43} \ergsec$, its age $t = 10^8 t_8 \year$, 
and the fraction $\fesc$ of its ionizing photons that escape
the galaxy to ionize the surrounding IGM (thereby becoming unavailable
for \lya\ line production) (RM01).  The result, conservatively ignoring recombination,  is
\begin{equation}
\rss \approx
{0.7 \pMpc \over 1+z}
\left\{ L_{43} t_8 \left( \fesc \over 1 - \fesc\right) \right\}^{1/3} ~~~.
\label{rsseq}
\end{equation}

Stellar population models for narrowband-selected \lya\ emitting
galaxies require $t_8 \la 1$ (and usually $t_8 \ll 1$) to produce the
observed range of \lya\ line equivalent widths (Malhotra \& Rhoads
2002, MR02).  The luminosity function for \lya\ galaxies at $z\approx 6$ 
shows $L_* \sim 10^{42.6} \ergsec$ (MR04), so 
that galaxies with $L_{43} \gg 1$ are rare.  Inserting
$t_8 \la 1$ and $L_{43} \la 1$  into equation~\ref{rsseq} and comparing the
result to equation~\ref{taueq}, we see that $\tau \ga 2$ at
line center even for reasonably bright \lya\ galaxies, and that
these galaxies' Stromgren spheres are not sufficient to prevent
\lya\ line suppression by an order of magnitude or more if the
surrounding IGM is neutral.

\paragraph{Velocity Offsets}
The observed \lya\ line is usually asymmetric (e.g. Rhoads et al 2003,
Dawson et al 2004) and offset to the red compared to other lines
(e.g., Shapley et al. 2003).  This offset is likely caused by
absorption or scattering of \lya\ photons in the blue wing of the
emission line in the interstellar medium of the emitting galaxy,
combined with gas motions (or outflows) in that medium. Resonant
scattering by the surrounding IGM will further suppress the blue wing
of the line, and accentuate the asymmetry.  The typical observed
velocity offsets for LBGs with \lya\ emission is $\sim 360 \kms$,
although the velocity offsets decrease for galaxies with higher
equivalent widths in the \lya\ emission (Shapley et al. 2003).
Such an offset can be incorporated in equation~\ref{taueq}
using $\Delta \lambda_{em} = \lambda_0 v_w/c$, where  $v_w$ (the ``wind'' velocity) denotes the offset of the \lya\ line relative to the systemic
velocity of the emitting galaxy.

If we describe the intrinsic profile as $f_\lambda(\lambda_0 +
\delta\lambda)$, where $\lambda_0 = 1215.67$\AA\ is the rest
\lya\ wavelength, then the unabsorbed line flux is $F_0 =
\int_{-\infty}^{\infty} f_\lambda(\lambda_0 + \delta \lambda) d(\delta
\lambda)$.  The absorbed line flux is $F = \int_{-\infty}^{\infty}
\exp[-\tau(\delta \lambda)] f_\lambda(\lambda_0 + \delta \lambda)
d(\delta \lambda)$, and the transmission factor is $T \equiv F / F_0$,
which is calculated by integrating over the line profile.

Santos (2004) has calculated the transmission factor for a large range
of model \lya\ line properties and IGM models His results show that $T
\la 1/3$ for a wide range of plausible models, even including those
with $v_w = 360 \kms$.  Models with $v_w = 0$ invariably show $T \ll
1/3$, unless the IGM is largly ionized.  Moreover, the variation of
$T$ with $v_w$ is fairly slow for $v_w > 360 \kms$; that is, a large
increase of $\langle v_w\rangle$ would be required to substantially
raise $\langle T \rangle$.

\paragraph{Galaxy clustering} 
Even though a single \lya\ galaxy does not produce a large enough
Stromgren sphere, there is a possibility that fainter galaxies
clustered around the observed \lya\ galaxy would contribute enough
photons to make the HII regions around the cluster reach the critical
size. This would require roughly 10-60 times more photons than
produced by a typical \lya\ galaxies observed (Wyithe \& Loeb 2005,
Haiman \& Cen 2005, Furlenatto et al. 2004) So far the evidence for
such clustering is mixed: Malhotra et al. 2005 see clustering in the
Hubble Ultra Deep field which was chosen to include a 25th magnitude
galaxy at z=5.8, so do Stiavelli et al 2005 around a SDSS quasar at
z=6.2; but Rhoads et al. (in prep) see no significant excess around an
actual z=6.5 \lya\ galaxy.

We here introduce a new version of the \lya\ reionization test wherein
aach observed \lya\ galaxy is seen because the IGM near that galaxy is
ionized in an otherwise neutral medium.  Each \lya\ galaxy thus
implies the presence of a certain volume of ionized IGM.  We combine
this ionized volume per source with the observed number density of
\lya\ galaxies to obtain a lower bound on the volume ionized fraction
of the IGM. {\it This limit is completely independent of whether the
  ionizing photons come from the \lya\ galaxy, its unseen neighbors,
  or any other source.}

\section{Ionized volume estimates}

\paragraph{Calculating the Transmission}
We need to calculate the \lya\ flux transmission expected
from a galaxy with a given Stromgren sphere radius, 
\lya\ line velocity offset, and IGM neutral fraction.
To do so, we must make some reasonable assumptions about the
line width and shape.
Observed \lya\ emission lines from high redshift galaxies are 
well described by a truncated Gaussian profile (Hu et al 2004,
Rhoads et al 2004):  
\begin{equation} 
f_\lambda(\lambda_0 + \delta\lambda) = \left\{
\begin{array}{ll} {2 F_0 \over \sqrt{2 \pi} \sigma}
\exp\left(-{(\delta \lambda)^2 \over 2 \sigma^2}\right)
& \hbox{if }\delta\lambda \ge 0 \\
0 & \hbox{if } \delta\lambda < 0 
\end{array} \right.
\end{equation}
where $\lambda_0 = 1215.67 (1+v_w / c)$\AA\ is the 
\lya\ central wavelength in the frame of the IGM surrounding the galaxy,
and the unabsorbed line flux is $F_0$.
The absorbed line flux is $F = \int_{-\infty}^{\infty}
\exp[-\tau(\delta \lambda)] f_\lambda(\lambda_0 + \delta \lambda)
d(\delta \lambda)$, and the transmission factor is $T \equiv F / F_0$.
We set  $\sigma=1.49$\AA, or rest
frame FWHM of $1.75$\AA\ (from the truncation point $\lambda_0$ to the
half-peak point on the red side of the line) -  chosen
to match typical observed high redshift \lya\ lines.

\begin{figure}
\epsscale{0.99}
\plotone{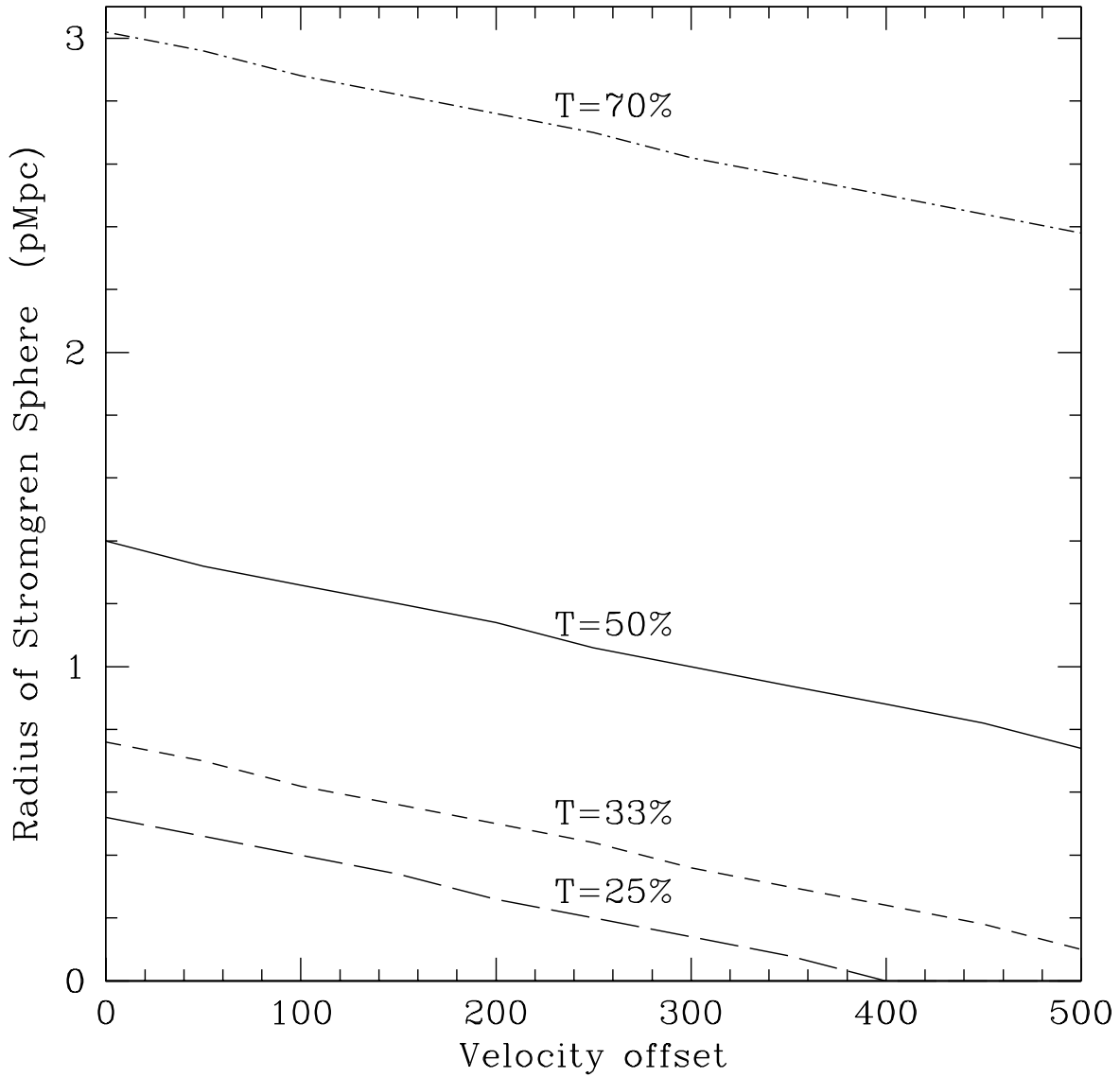}
\caption{This figure shows the minimum radius of the ionized bubble needed
for 70\% (dot dashed line), 50\% (solid line), 33\% (short dashed line) or 25\% (long dashed line) of \lya\ line flux to escape as a function of the velocity offset between the \lya\ emission line and the intrinsic velocity of the galaxy. 
Currently, a transmission factor of less than 50\% is ruled out by MR04, and
the best guess for velocity offset is $\approx 350 \kms$; which would imply $R_{ss}=0.93$ physical or 7 comoving Mpc.}
\label{crit_rad}
\end{figure}

We determined $T$ by numerical integration for a grid of
$v_w$ and $\rss$.  We then inverted the result to determine the
Stromgren sphere radii corresponding to 25\%,
33\%, 50\%, and 70\% transmission for a range of \lya\ velocity
offsets $v_w$.  The result is summarized in figure~\ref{crit_rad}.

\paragraph{The Minimum Ionized Fraction: Analytic results}
\label{analytic}
We next combine our calculated minimum Stromgren sphere radii with
the observed number density of $z\approx 6.5$ \lya\ emitting galaxies
to place a lower bound on the volume ionized fraction of the intergalactic
medium.  The problem is closely related to the void probability function,
or to a counts-in-cells formalism.  The volume neutral fraction 
$\xvn \equiv 1 - \xvion$ can be found by calculating the probability that
an arbitrary point in space lies in a void of radius $\ge \rss$ in
the \lya\ galaxy distribution,
or (equivalently) by calculating the probability that a randomly
placed, spherical cell of radius $\rss$ will contain no \lya\ galaxies.

The zero order estimate of the ionized volume fraction is
simply $\xvion \approx \ff \equiv n \vss$.  Here $n$ is the 
volume number density of the detected \lya\ emitting galaxies, 
$\vss = (4 \pi / 3) \rss^3$ is the volume of
the minimum size Stromgren sphere that renders them visible,
and $\ff$ is the filling factor of these spheres.  
The estimate $\xvion \approx \ff$ is accurate for $\xvion \ll 1$.  
At larger ionized fractions, we need to correct for overlap of the Stromgren
spheres.  

If the locations of the \lya\ galaxies are uncorrelated, the 
residual neutral volume fraction becomes $\xvn = \exp(-\ff)$
and the corresponding ionized fraction is $\xvion = 1 - \exp(-\ff)
= \sum_{j=1}^\infty (-1)^{j-1} \ff^j / j!$.

The degree of overlap will be further enhanced by spatial 
correlations among galaxies.  To account for this overlap, we
can assume that the observed \lya\ emitters are drawn from
a distribution with two-point angular correlation function
$\xi(r) \approx (r/r_0)^{-\gamma}$.  We take values $r_0 \approx 4\cMpc$
and $\gamma = 1.8$, which are typical of high redshift galaxy populations
and in particular are consistent with observed \lya\ galaxy
populations (e.g., Ouchi et al. 2003, Kovac et al. in prep.).
This correlation function implies that the mean number of neighbors
within distance $r$ of an arbitrarily selected
galaxy should be $\langle N_p\rangle = (4\pi/3) r^3 n 
\left( 1 + 3 (r/r_0)^{-\gamma} / (3-\gamma) \right)$ (Peebles 1980,
eq.~31.8).  Given a pair of ionized spheres with radius $\rss$,
whose centers are separated by distance $r$ (with $0\le r/\rss \le 2$),
the volume of their overlap region is $\vss \times \left( 1 - 0.75 
(r / \rss)  + (r/\rss)^3/16 \right)$.  
We integrate the expected volume ``lost'' to overlap within 
a survey by combining the two foregoing results:
\begin{equation}
\begin{array}{lll}
\xvion & = 
 & \ff - {1 \over 2}\ff^2
  \left\{ 1 + 3\left(\rss \over r_0\right)^{-\gamma}
  \left[ {2^{3-\gamma} \over 3-\gamma}
         - {3\over 4}{2^{4-\gamma} \over 4-\gamma}
         {2^{6-\gamma} \over 16(6-\gamma)} \right] \right\}
\end{array}
\label{order2}
\end{equation}
While this calculation accounts for correlations, it omits
third order and higher terms in $\ff$.
The coefficient for the $\ff^3$ term would depend on the three
point correlation function.  Thus
equation~\ref{order2} will tend to {\it underestimate\/}
the volume ionized fraction, and is conservative.

\paragraph{Minimum Ionized Fraction: Numerical simulations}
To achieve accurate results at higher source densities,
we simulated correlated distributions of galaxies and
directly calculated the volume fraction enclosed by the
union of their Stromgren spheres.  We performed these
calculations for a range of correlation lengths, 
Stromgren sphere sizes, and source densitites.
We generated the set of correlated $(x,y,z)$ triples
using a Mandelbrot-Levy random walk prescription 
(see Peebles 1980, section~62).  A power law distribution of
step lengths $s$, $p(s) \propto s^{-1.2}$,
ensures a power law correlation function with
the desired slope $\gamma = -1.8$.  The correlation
length $r_0$ is related to the minimum step length (which is
always $\ll \rss$ and $\ll r_0$) and to the number
of independent random chains occupying the sample volume.
To tune the correlation length to a desired value, we
vary the number of chains used and also (for small correlation
lengths) add a suitable number of entirely uncorrelated
$(x,y,z)$ triples.

We verified that this prescription generated the desired two-point
correlation properties.
The volume used for the simulations was
$\sim 10^6 \cMpc^3$, and the volume fraction was calculated on a grid
with cell size $= \rss/2$.  (A smaller cell size does not
significantly change the results) Finally, we verified that the
simulations reproduce the analytic results for an uncorrelated
distribution at any filling factor, and for a correlated distribution
at $\ff \la 1$ (where the $\ff^3$ and higher terms are small). 

\begin{figure}
\epsscale{0.99}
\plotone{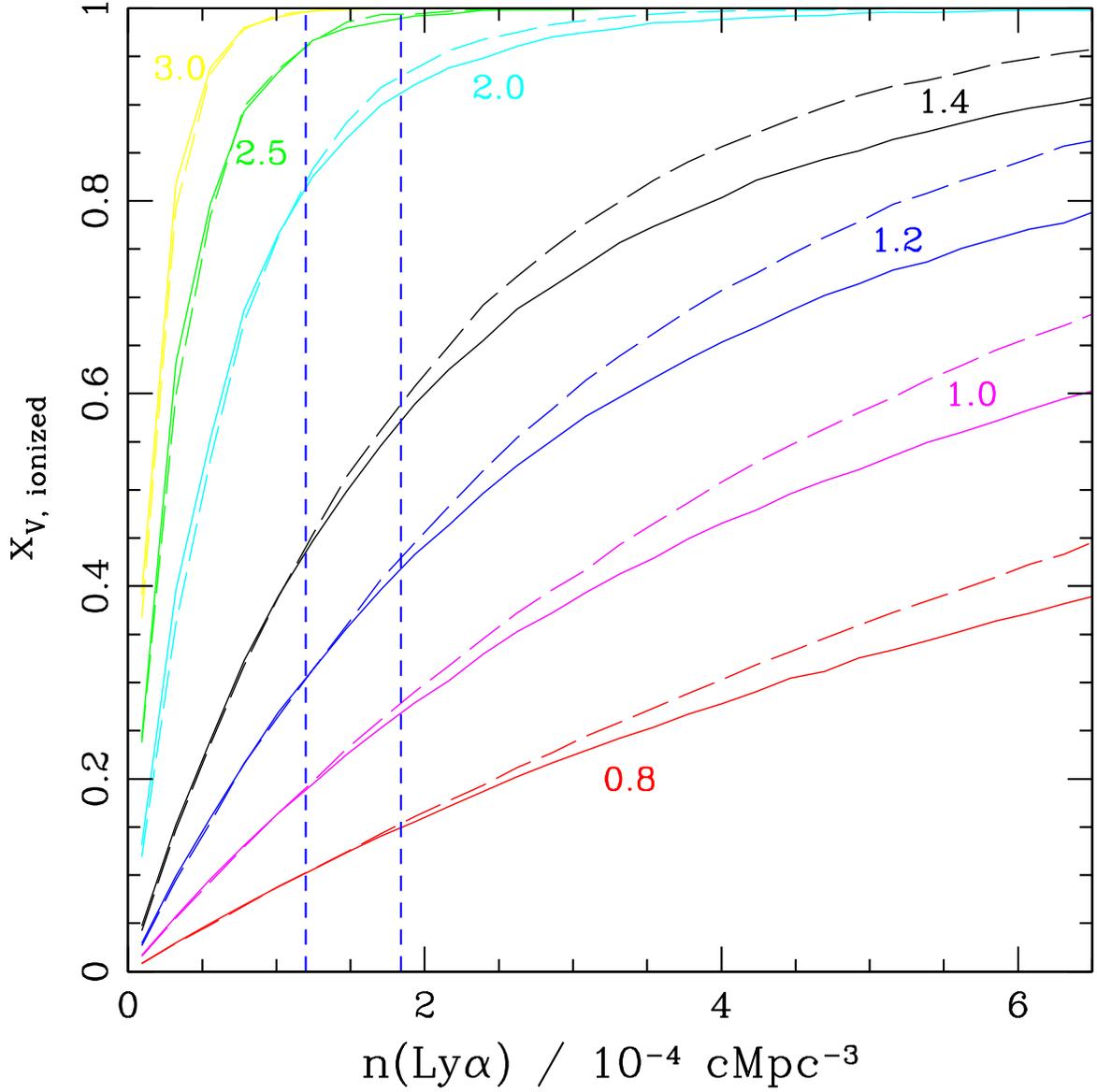}
\caption{The volume ionized fraction as a function of observed
  \lya\ source density is plotted for a range of minimum Stromgren
  sphere sizes and correlation lengths. From highest to lowest, these
  sets of curves correspond to $\rss = 3.0\pMpc$ (yellow), $2.5 \pMpc$
  (red), $2.0 \pMpc$ (magenta), $1.4 \pMpc$ (green), $1.2 \pMpc$
  (cyan), $1.0 \pMpc$ (black) and $0.8 \pMpc$ (blue). For each set of 
  curves, the line style indicates the correlation length: $r_0 = 0$ (dotted),
  and $4 \cMpc$ (solid).
  Vertical lines at $1.2\times 10^{-4}$ and $1.8\times 10^{-4}
  \cMpc^{-3}$ mark the range of \lya\ number density observed by Taniguchi et al
  (2004).  For the best-guess parameters $1 \la \rss \le 1.4 \pMpc$
  and $r_0 \approx 4 \cMpc$, the Taniguchi data imply a lower bound of
  $0.2$ to $0.5$ on $\xvion$.}
\label{simfig}
\end{figure}

Results
of the simulations are shown in figure~\ref{simfig}, where we show
curves of $\xvion$ as a function of the \lya\ galaxy number density
for a plausible range of $\rss$ and $r_0$ values.
Were we to plot $\xvion$ as a function of $\ff$ (rather than $n$),
we would see that  $\xvion$ is a function of just two
variables, $\ff$ and the ratio $(r_0 / \rss)$, as one would expect from
the discussion in section 2.

\paragraph{Effect of a Partly Ionized Ambient Medium}

\begin{figure}
\epsscale{0.99}
\plotone{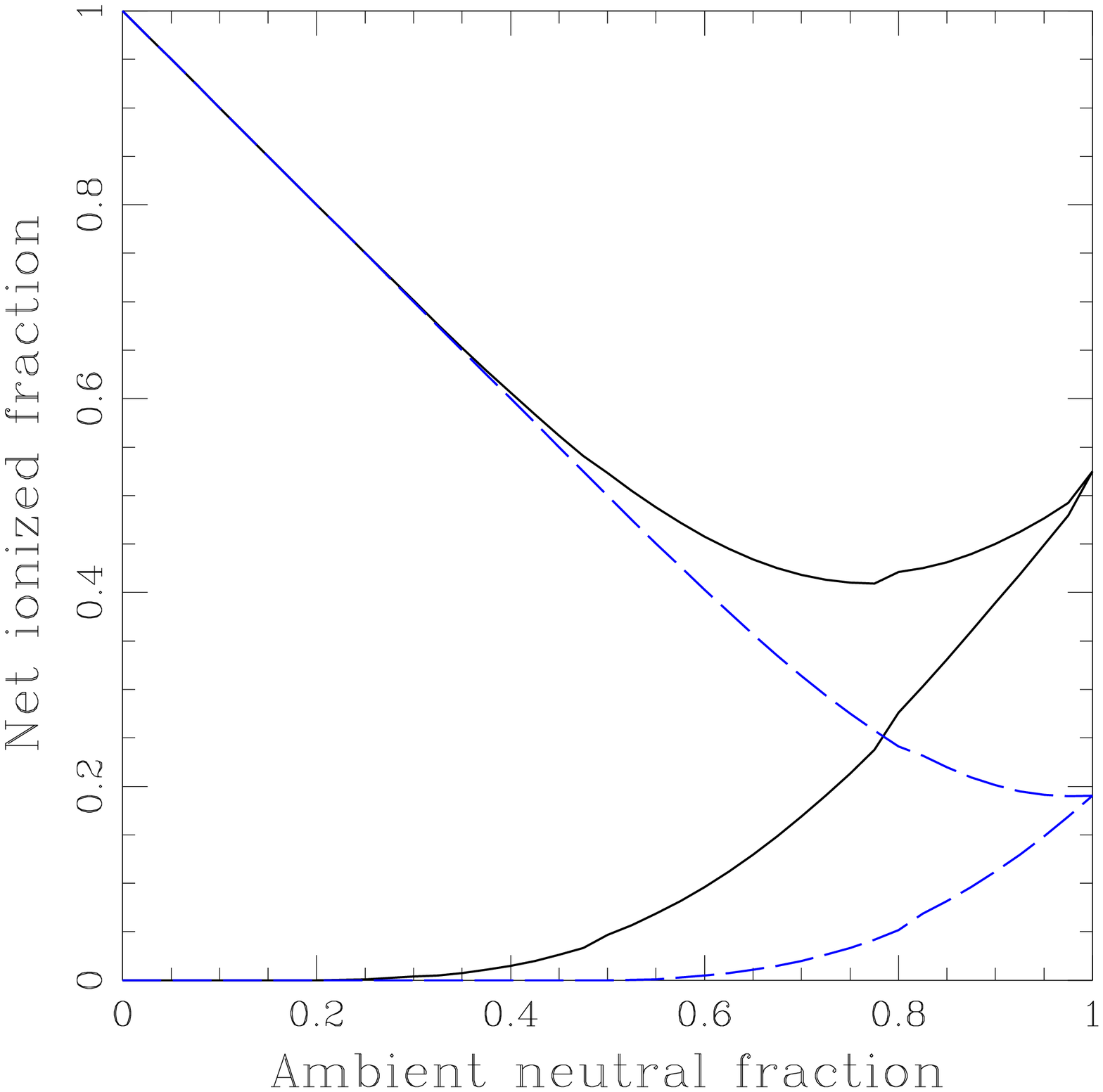}
\caption{The ionized gas fraction in the IGM is plotted as a function
of the neutral fraction in the ambient medium (outside Stromgren spheres).
We have assumed a number density of $1.5\times 10^{-4} \cMpc^{-3}$
(Taniguchi et al 2004).  Two solid (black) curves show the result for
$v_w = 0$.  The lower curve shows the amount of ionized gas enclosed in
Stromgren spheres around \lya\ galaxies, and the upper curve shows the
total ionized gas fraction (including that in the ambient medium).
The dashed (blue) curves show the same quantities for $v_w = 360 \kms$.}
\label{ambfig}

\end{figure}

Up to now, we have considered Stromgren spheres embedded in a
fully neutral ambient intergalactic medium.
If this medium is already partially ionized outside the bubbles,
then \lya\ flux is transmitted more easily, and the Stromgren spheres
required to achieve a particular \lya\ transmission
are smaller.  However, this scenario requires a large amount of ionized
gas mixed into the ambient medium.
Figure \ref{ambfig} shows the net ionized fraction
of ionized gas (counting the Stromgren spheres + ambient ionized gas) as
a function of ambient neutral fraction for  $v_w = 0 $ and $360 \kms$

\section{Summary and Future outlook}
Observations of \lya\ emitters provide a powerful method to 
probe the neutral fraction of the IGM and the volume of the ionized
gas. In an earlier paper we explored the effect on the \lya\ luminosity 
function and constrained the average neutral fraction
of the IGM to be $\langle {x_{\hbox{\scriptsize n}}} \rangle < 30$--$50\%$ 
(MR04, see also Stern et al. 2005). Further 
investigations by Haiman \& Oh 2005 and Furlennato et al. 2005 agreed 
with these estimates. In this letter we place a lower limit on the volume in 
ionized IGM $\xvion = 20-50 \%$ for best known parameters of transmission factor $T = 0.5$, source density $n(ly \alpha)=10^{-4}$, and $v_w=360 - 0 \kms$.

With better observations, these limits can be substantially
improved. Currently, the largest \lya\ galaxy sample at z=6.5 comes
from Subaru observations (Taniguchi et al. 2005) which reach just
about $\L_*$. Going deeper is then expected to yield a higher density
of \lya\ emitters. If the current observed number density of sources
could be increased by a factor of 2 or 3 with deeper observations, the
lower limit to $\xvion$ would increase by a factor of 2. Bigger
samples (both at z=6.5 and z=5.7) would also improve the estimate of
the attenuation factor. MR04 concluded that attenuation of a factor
of two was marginally allowed by present data. If we could reduce this limit to
1.4, i.e a transmission factor of 70\%, we would conclude that the
minimum Stromgren sphere radius needed would be 3 \pMpc\ for offset
velocity of zero, and 2.6 \pMpc\ for a velocity offset of 350
\kms\ (Figure 1). In that case the Taniguchi et al. 2005 data would
imply nearly $\xvion =100\%$.  Moreover, we should not forget that we
are using the smallest value of $R_{SS}$ that allows the transmission of
the \lya\ line. The ionized bubbles could certainly be larger.

 Measurement of the velocity offset between
\lya\ redshift and systematic velocity of the galaxy will allow us to
further refine the Stromgren sphere radius (see Figure 1). Currently
our estimates come from such measurements on Lyman break 
galaxies (LBGs), for which the higher the
EW of the \lya\ line, the smaller the velocity offset (Shapley et
al. 2003).  But the overlap between the LBG samples and the \lya\ samples
is minimal; only the LBGs with highest EW (EW $ \approx 50 $ \AA) would
even be selected by the \lya\ searches using Narrow-bands. So it
is plausible that the velocity offsets for the \lya\ emitters (median EW
$\approx 100$\AA;  (MR02, Dawson et al. 2004) would
be even lower. 

Spectroscopy of large, complete samples would also yield the
information on their distribution and clustering in the line-of-sight
dimension, which can be used to determine the overlap in the ionized
bubble around the galaxies. Then we can go beyond the zeroth order
estimate of $ \xvion $ to learn about the topology of the ionized
bubbles and thus determine whether substantial overlap has happened
(Rhoads, in prep).  Pre-overlap phases of reionization can also be detected
by heightened correlation between \lya\ sources (Furlanetto et al. 2005).
If the sources required for ionizing photons are strongly clustered
on $\approx 100 cMpc$ scales as observed (Malhotra et al. 2005) or
predicted (Cen 2005), we should see dramatic field-to-field variations 
in \lya\ number density prior to reionization. 

These tests only require the detection of statistically useful samples
of \lya\ emitters and are  eminently practical with present technology
at optical wavelengths, and their extension to redshifts $z>7$ through
near-infrared observations is close at hand.

\acknowledgements 
We thank Michael Strauss for useful discussions.


\end{document}